\documentclass{pic2012}
\def\runauthor{Stefano Argir\`o}
\def\shorttitle{Bottom Production, Spectroscopy and Lifetimes}
\begin{document}

\title{Bottom Production, Spectroscopy and Lifetimes
}

\author{Stefano Argir\`o\\
on behalf of the CMS, ATLAS, CDF, D0,\\ 
H1, LHCb and Zeus Collaborations}

\address{University of Torino\\
Physics Department\\
via Pietro Giuria 1, Torino, Italy\\
E-mail: stefano.argiro@unito.it}

\maketitle

\abstracts{In this contribution, I give a brief overview of the latest results related to the production, spectroscopy and lifetimes of bottom and charm hadrons. Several interesting experimental results were presented in this field in 2012. The focus will be on the findings of experiments performed at hadron colliders, since the LHC is the main player this year, with a brief mention about electron-proton collider results.  
}


\section{Introduction} 
The phenomenology of heavy-flavored (HF) hadrons is an essential workbench for the understanding of strong interactions. Measurements in this field can provide useful tests of QCD, and give input for tuning models, and refining event generators. 
At hadron colliders of TeV energies, heavy quarks are produced copiously, and the amount of data that can be stored for these studies are limited only by the allocated data-logging bandwidth and by practical constraints connected to the storage and reprocessing of the large volumes of information. 

The start of the LHC opened a new era in this field, extending the reach in transverse momentum and introducing into the arena, for the first time at a hadron collider, a detector completely dedicated to heavy-flavor physics in the forward region (LHCb). The general-purpose LHC detectors, ATLAS and CMS, are also adding high-quality data and complement well the rapidity and transverse momentum coverage of LHCb. In addition, the now ended Tevatron experiments CDF and D0 are still producing new results from the analysis of their datasets, as are H1 and ZEUS at HERA.

In this contribution, I will review in particular three aspects of heavy-flavor physics: production, spectroscopy and lifetimes. Other important topics, like CP-violation, are discussed elsewhere in these proceedings \cite{pic2012:cp}.
The area of heavy-flavor production is the one in which the largest quantity of results was presented in the past twelve months. Since it would be impossible to present them all, I've tried to arrange them by experimental technique and to provide a few examples of measurements executed with  each of these techniques. 
There where several surprises in the area of spectroscopy, with a number of new states and decay modes being discovered.  The precision measurement of the lifetime of heavy-flavored hadrons has two important highlights: new measurements that  shed light on the $\Lambda_b$ lifetime puzzle and the precision measurement of the $B_s$ lifetime.
I hope that the following  pages will provide a useful compendium of heavy-flavor results published this year.

\section{Heavy-Flavor Production}
 
A precise understanding of the mechanisms of HF production can be viewed as one of the ultimate tests of QCD, and allows us to probe our knowledge of the fundamental constituents of matter and their interactions. Several ingredients are in fact needed to predict HF production cross sections. First, a precise knowledge of the structure of the projectiles, protons or anti-protons, must be available in order to make predictions about the final state. This translates into the need to know the parton distribution functions (PDF) of the nucleons in the kinematical regions of interest. For this piece of the puzzle, the high-energy community has profited from a decade of measurements carried out at ZEUS and H1 using the HERA electron-proton collider at DESY.
  
The next step is the calculation of the so-called \emph{hard process}, i.e., the hard scattering or annihilation in which the partons inside the projectiles participate. This is a perturbative process that, in principle, can be calculated with the desired precision. However, the calculation is not free of technical difficulties, for example, including  the higher-order perturbative diagrams and applying a consistent renormalization scheme in order  to re-absorb the infrared divergences originating from the emission of soft gluons. 

Once the perturbative hard process has been consistently accounted for, one faces the most difficult part of the calculation, which is the problem of describing the fragmentation of the partons that  are the products of the hard process itself. Fragmentation describes how heavy-flavored hadrons are produced from the heavy quarks that are involved in the hard scattering. This is a inherently nonperturbative process. Current models use either fits to $e^+e^-$ collider data or parton-shower Monte Carlo simulation techniques. Finally, in order to include the experimentally accessible final states, the weak decay of the bottom (or charm) hadrons must be  properly taken into account.

\begin{figure}[h]
\begin{center} 
    \includegraphics[width=0.9\textwidth]{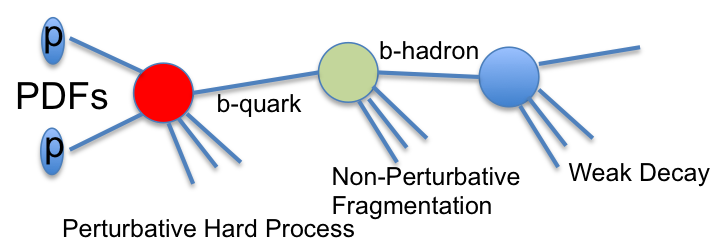}
\caption{Pictorial representation of the production process of heavy-flavored hadrons.}
\label{fig:hfprod}
\end{center}
\end{figure}

The process of heavy-flavor production and decay is illustrated in Fig. \ref{fig:hfprod} (inspired by \cite{cacciari}), where the red circle represents the hard interaction, the green one the fragmentation process and the blue the weak decay. 
As a concrete example, suppose we want to  predict  the cross section for the process $pp \rightarrow B + X \rightarrow J/\psi + Y$. Following the above scheme, we can write :

\[
\frac{d\sigma (b\rightarrow B \rightarrow J/\psi)}{dp_{T}} =
   \frac{d\sigma(b)}{d \hat{p}_T} \otimes
   f(b \rightarrow B) \otimes 
   g(B \rightarrow J/\psi),
\]

where $p_T$ is the transverse momentum of the $J/\psi$, $\hat{p}_T$ is the transverse momentum of the parton undergoing the hard scattering (described by the PDFs), $f$ is the fragmentation function and $g$ describes the weak decay.

Several kinds of experimental measurements can be used to probe our understanding of HF production. One of the easiest ways is by measuring electrons and muons originating from the decay of HF hadrons. Another possibility is to perform an inclusive measurement, using the  signature provided by a displaced $J/\psi$ to tag the presence of a b hadron. Heavy flavors can also be studied in association with jets, for example by measuring  $D^*$ mesons produced in jets. HF production can also be studied via exclusive final states ($B^+$, $B^0$, $B_s$, $\Lambda_b$). Quarkonium and bottomonium are examples of systems that can be investigated in a relatively  easy way, and can give valuable information about the dynamics of the strong interaction. In the following sections, I will give some examples of recent measurements of the kinds listed above. 
Production of HF in association with vector bosons and multiple-HF production (e.g., double $J/\psi$) are covered by the contributions at this conference of Bob Hirosky and Ellie Dobson, respectively. 

\subsection{Inclusive cross section measurements of electrons and muons from HF decays}

\begin{figure}
\begin{center} 
    \includegraphics[width=0.42\textwidth]{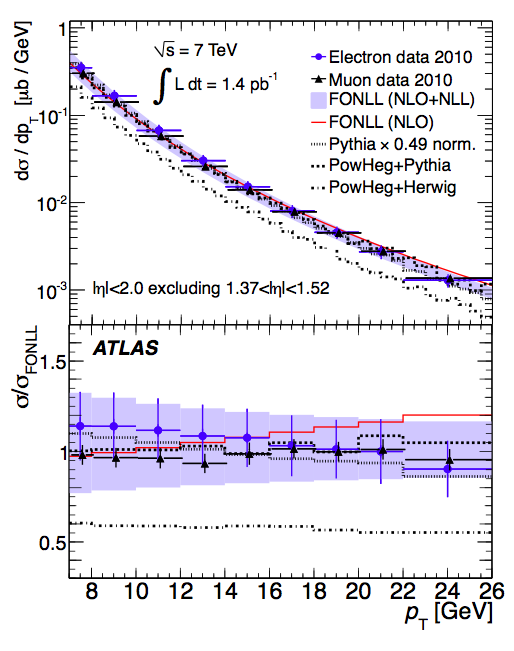}
    \includegraphics[width=0.52\textwidth]{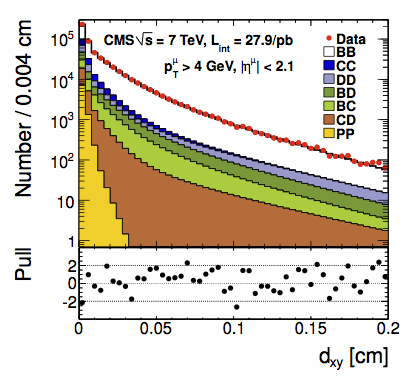}
\caption[*]{Left: inclusive electron and muon cross sections as a function of $p_T$ from heavy-flavored hadrons by ATLAS~\cite{atlas:emu}. Right: distribution of $d_{xy}$ for muons pairs from heavy-flavor decays from CMS~\cite{cms:bbarmumu}. The different components are combinations of muons originating from a b quark (B), c quark (C), prompt tracks (P) and decays in-flight (D). In both figures, the bottom plots represent the comparison of theoretical predictions to the data.}
\label{fig:a}
\end{center}
\end{figure}

Both ATLAS and CMS have made inclusive measurements of the HF production cross section using  electrons and muons. ATLAS \protect\cite{atlas:emu} presented an inclusive cross section measurement for both electrons and muons, after subtraction of the contributions from W, Z and Drell-Yan processes. The data are in agreement with predictions from FONLL~\cite{fonll} and POWHEG~\cite{powheg}, as can be seen in Fig. \ref{fig:a} (left). CMS studied the production cross section of muon pairs from the process $pp \rightarrow b \overline{b}X \rightarrow \mu \mu X'$ \protect\cite{cms:bbarmumu}. Muons from b decays are selected by using the decay length  in the transverse plane to discriminate against promptly produced muons. Data are compared to MC@NLO~\cite{mcnlo} predictions and are found to be compatible within the experimental and theoretical uncertainties (Fig. \ref{fig:a}, right).

\subsection{Inclusive b production cross sections via $H_b \rightarrow D^{*+} \mu^- X$}
\begin{figure}
\begin{center} 
    \includegraphics[width=0.45\textwidth]{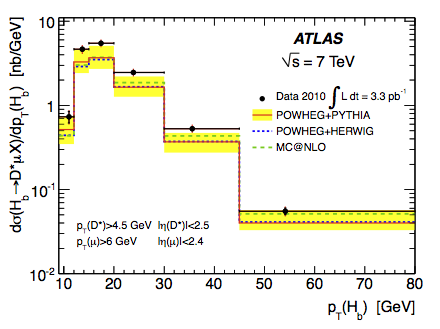}
    \includegraphics[width=0.45\textwidth]{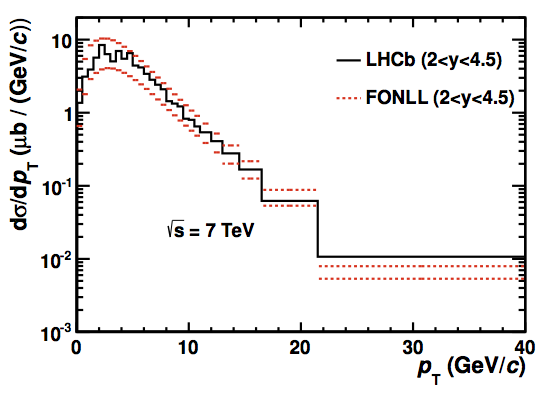}
\caption[*]{Left: cross section for the process $H_b \rightarrow D^{*+} \mu^- X$ from ATLAS~\cite{atlas:dstar}. Right: exclusive $B^+$ production cross section from LHCb. Both figures show the experimental data as well as theoretical predictions~\cite{lhcb:b}.}
\label{fig:b}
\end{center}
\end{figure}

ATLAS presented a measurement of $H_b \rightarrow D^{*+} \mu^- X$, with $D^{*+} \rightarrow \pi^+ D^0$ and $D^0 \rightarrow K^- \pi^+$ (and charge conjugates) \cite{atlas:dstar}. Here, $H_b$ is a generic hadron containing a b quark. The $D^{*+}$ candidate is required to have $|\eta| < 2.5$ (where $\eta$ represents pseudorapidity) and $p_T>4.5$ GeV. Monte Carlo simulations are used to subtract nondirect semileptonic processes that contribute to the $D^{*+} \mu^-$ data sample. Results are compared to POWHEG and MC@NLO, and found to be systematically higher than predictions, although not by a lot considering the theoretical uncertainties, as can be seen in Fig. \ref{fig:b} (left).

\subsection{Exclusive $B^+$ production cross section at 7 TeV}

The LHCb Collaboration has published a measurement of the exclusive cross section for $pp \rightarrow B^\pm \rightarrow J/\psi K^+$ with 35 nb$^{-1}$ of data at $\sqrt{s}$ = 7 TeV \cite{lhcb:b}. This is the first measurement of B production at hadron colliders in the forward region ($2.0<y<4.5$, where $y$ represents rapidity). The transverse momentum of the B hadron reaches up to 40 GeV. Good agreement with FONLL predictions is found. The $B^+$ production cross section as a function of $p_T$ can be seen in Fig. \ref{fig:b} (right). The uncertainties in the theoretical calculation arise primarily from  the choice of renormalization and factorization scales and from the b-quark mass. 

\subsection{Exclusive $\Lambda_b$ production cross section}

\begin{figure}
\begin{center} 
    \includegraphics[width=0.45\textwidth]{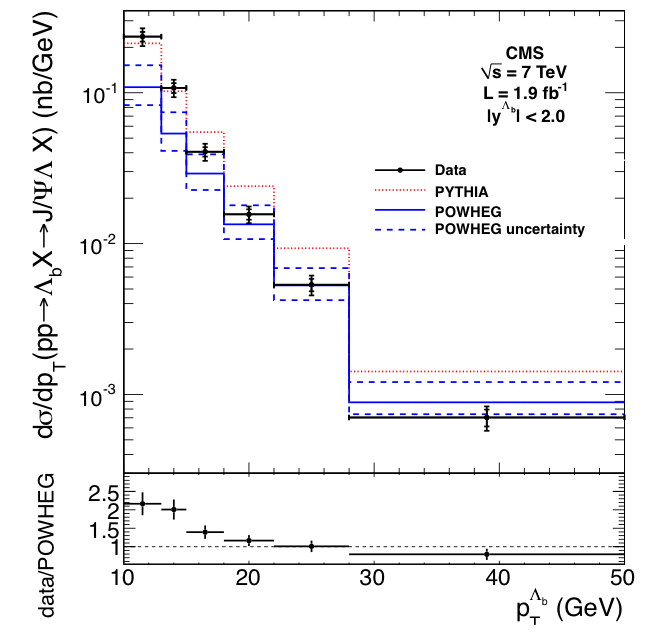}
    \includegraphics[width=0.45\textwidth]{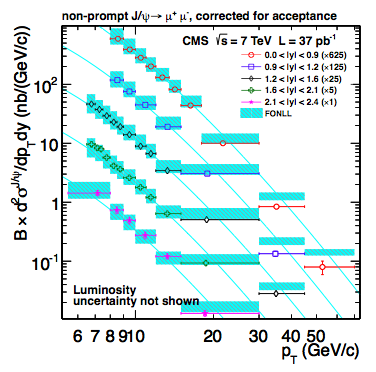}
\caption[*]{Left: $\Lambda_b$ production cross section as a function of $p_T$ from CMS~\cite{cms:lambdab}, compared to several theoretical predictions. Right: production cross section of nonprompt $J/\psi$ from CMS~\cite{cms:jpsipsip}, in different rapidity ranges, compared to FONLL predictions.}
\label{fig:c}
\end{center}
\end{figure}

The CMS Collaboration has published a study of the $\Lambda_b$ differential cross section, as a function of transverse momentum and rapidity, in the channel $\Lambda_b \rightarrow J/\psi \Lambda^0 \rightarrow \mu \mu p \pi^-$, using a displaced-$J/\psi$ trigger \cite{cms:lambdab}. The transverse momentum of the $\Lambda_b$ ranges from 10 to 50 GeV and the rapidity range is $|y|<2.0$. In general, good agreement is found with theoretical predictions from POWHEG, but the transverse momentum spectrum is significantly steeper than expected, as can be observed in Fig. \ref{fig:c} (left). The spectrum also falls more rapidly than the spectrum of B mesons. 
 The CMS Collaboration has also measured, in the same paper, the ratio $\sigma(\overline{\Lambda}_b) / \sigma(\Lambda_b)$ finding no significant deviation from unity, although with relatively large uncertainties.

\subsection{Inclusive $J/\psi$ production}

Valuable information about bottom production can be extracted from the inclusive study of events containing a $J/\psi$. A now well-established experimental technique consists in a two-dimensional maximum-likelihood fit to the mass and lifetime, allowing one to distinguish between promptly produced $J/\psi$ and $J/\psi$ originating from decays of b hadrons. CMS has presented the latest measurement in \cite{cms:jpsipsip}. Agreement with NRQCD predictions is found for the prompt component, and with FONLL for the nonprompt component. For example, in Fig. \ref{fig:c} (right) the nonprompt $J/\psi$ production cross section  for several rapidity ranges, as measured by CMS as a function of $p_T$, is compared to FONLL predictions. There is also very good consistency of the results from CMS, ATLAS \cite{atlas:jpsi} and CDF \cite{cdf:jpsi}.

\subsection{$\Upsilon$ production and polarization}

\begin{figure}
\begin{center} 
    \includegraphics[width=0.54\textwidth]{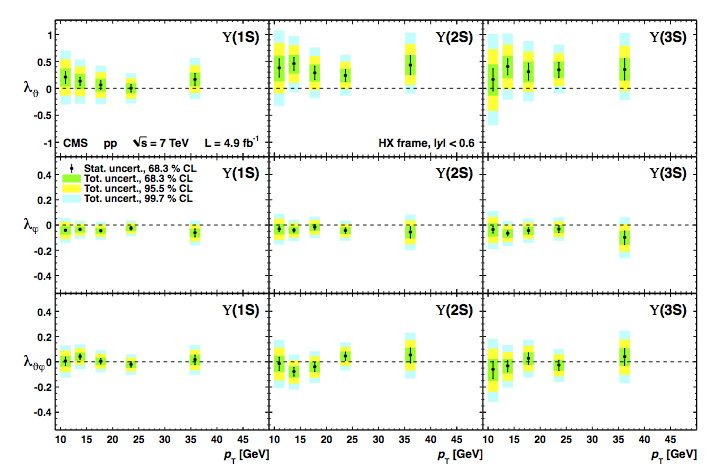}
    \includegraphics[width=0.41\textwidth]{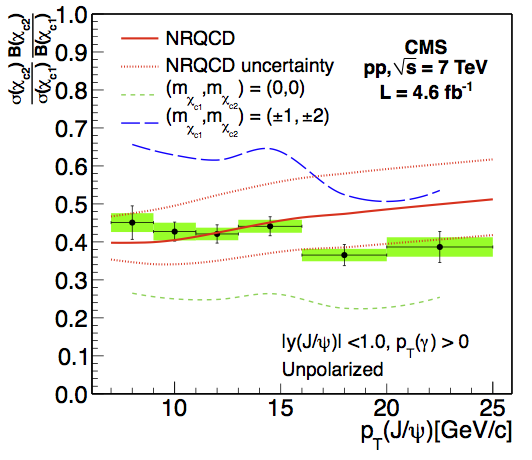}
\caption[*]{Left: polarization parameters of the $\Upsilon(1S)$, $\Upsilon(2S)$, and $\Upsilon(3S)$ from CMS~\cite{cms:ypol}. Right: measurement of the relative prompt-production cross section of $\chi_{c2}$ and $\chi_{c1}$ from CMS, compared to the NRQCD prediction~\cite{cms:chi2chi1}. }

\label{fig:d}
\end{center}
\end{figure}

In the quarkonium sector, the CMS Collaboration has presented new results on the production cross section \cite{cms:yprod} and polarization \cite{cms:ypol} of the $\Upsilon(1S), \Upsilon(2S)$, and  $\Upsilon(3S)$. Cross sections are measured differentially as a function of $p_T(\Upsilon)$ up to 35 GeV, and as a function of rapidity up to $y(\Upsilon)= 2.4$.  Cross section measurements are in agreement with NRQCD predictions. Polarization measurements performed up to $p_T(\Upsilon)=50$ GeV confirm the surprising result from CDF \cite{cdf:ypol} that all three states seem to be produced unpolarized, contrary to expectations. The result is illustrated in Fig. \ref{fig:d} (left). The quantities $\lambda_\theta$, $\lambda_\phi$ and $\lambda_{\theta,\phi}$ are coefficients of the $\sin^2 \theta $, $\sin^2 \theta  \cos 2\phi $ and $\sin 2\theta \cos \phi$ terms, respectively, in the expression for the angular distribution of the $\Upsilon \rightarrow \mu \mu$ decay. 

\subsection{$\chi_{c2}$/$\chi_{c1}$ prompt-production ratio}
The CMS Collaboration has recently presented a precise measurement of the prompt-production ratio of $\chi_{c2}$ and $\chi_{c1}$ \cite{cms:chi2chi1}. In this analysis, the $\chi_c$ mesons are detected using the radiative decay $\chi_{cJ} \rightarrow J/\psi + \gamma$, where the photon is detected using its conversion into an $e^+e^-$ pair in the silicon tracker. The measurement extends up to $p_T(J/\psi)$= 25 GeV. This technique gives a mass resolution of about 6 MeV, permitting a clear separation  between the two states. The experimental results are compared to NRQCD and $k_T$-factorization predictions~\cite{kt}. The NRQCD prediction is compatible with the data within certain assumptions on the polarization of these states, as can be seen in Fig. \ref{fig:d} (right).

\subsection{HF photoproduction at HERA}

The ZEUS and H1 experiments at HERA explored heavy-flavor photoproduction in electron-proton collisions. Several measurements were presented in 2012. Boson-gluon fusion is the dominant production mechanism in these experiments. To mention a few examples, H1 has studied charm photoproduction with $D^*$ and D jets \cite{h1:charm} and beauty photoproduction via the process $ep \rightarrow e b \overline{b} X \rightarrow eee X'$ \cite{h1:b}. The experimental results are in agreement with theoretical predictions in both cases.

\section{Spectroscopy}

Spectroscopy has historically been an important tool for progress in physics. We can think, for example, of how the study of atomic transitions helped in understanding quantum mechanics and electrodynamics. Mesons and baryons containing a b quark can respectively be seen as the hydrogen  and helium atoms of QCD, and therefore can play an important role in the understanding of strong interactions. The spectra of b hadrons are described in the framework of heavy-quark effective theory \cite{hqet}. In this framework, the heavy quark is viewed as a static color source,  and its spin is taken to be decoupled  from the system. As an example, we can consider the $b\overline{u}$ case. The system is characterized by three quantum numbers: $L$, the orbital angular momentum of the system, $j_q$, the angular momentum of the light quark, and $J$, the total angular momentum of the system. For $L=0$ the theory predicts the B meson and the radially excited $B^\prime$ when J=0, and the B$^*$ and $B^{* \prime}$ when J=1. For $L=1$ we have two doublets, one with $j_q= 1/2$ and the other with $j_q=3/2$. In the first case, we have the B$^*_0$ with $J=0$ and the B$^*_1$ with $J=1$. In the second case, the $B_1$ with $J=1$ and B$^*_2$ with $J=2$.  The four states are collectively called B$^{**}$. The $B_0^*$ and $B_1^*$ are expected to decay via an S wave and, therefore, are expected to be wide resonances, while the $B_1$ and $B^{*}_2$ should decay via D wave and be narrow.

\begin{figure}
\begin{center} 
    \includegraphics[width=0.40\textwidth]{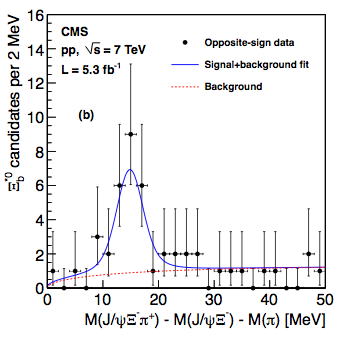}
    \includegraphics[width=0.55\textwidth]{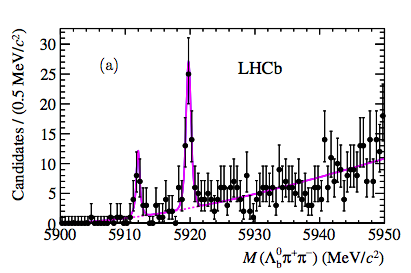}
\caption[*]{Left: the $\Xi_b^{0*}$ signal published by CMS~\cite{cms:xib}. Right: the $\Lambda_b^{0*}$ discovered by LHCb~\cite{lhcb:lambdabstar}.}
\label{fig:e}
\end{center}
\end{figure}

\subsection{Excited B mesons}
While the B$_1^0$ and B$_2^*$ were observed at the Tevatron \cite{cdf:bstar}, their charged counterparts have not been previously seen. The LHCb Collaboration reported the observation of the B$_1^+$ and B$^{*+}_2$ through the decays $B_1^+ \rightarrow B^{0*} \pi^+$ and $B^{+*}_2 \rightarrow B^0 \pi^+ $ \cite{lhcb:bstar}. The B$^{0*}$ and B$^0$ mesons are reconstructed using  several decay modes: $J/\psi(\mu^+ \mu^-) K^*(892)^0(K^+ \pi^-) $, $D^+(K^+ \pi^+ \pi^-) \pi^+$ and  $D^+( K^+ \pi^+ \pi^-) \pi^+ \pi^- \pi^+$. The $B_1^+$ mass is measured to be  5726.3  $\pm$ 1.9 (stat.) $\pm$ 3.0 (syst.) MeV  with a significance of about 10$\sigma$, while the $B^{*+}_2$ is found at 5739.0  $\pm$ 3.3 (stat.) $\pm$ 1.6 (syst.) MeV with a significance of 4$\sigma$. The masses are in good agreement with theoretical predictions.

\subsection{Excited B baryons}

The LHCb Collaboration has made the first observation of the $\Lambda_b^{0*}$, an orbitally excited $bud$ system \cite{lhcb:lambdabstar}. Two narrow states are observed in the $\Lambda_b^0\pi \pi$ invariant-mass spectrum, using the decays $\Lambda_b^0 \rightarrow \Lambda_c^+ \pi$ and $\Lambda_c^+ \rightarrow pK \pi$, as can be seen in Fig. \ref{fig:e} (right). Orbitally excited states are expected for J = 1/2 and J = 3/2. The higher mass state was confirmed by CDF \cite{cdf:lambdabstar}. 

CMS has found  $>5\sigma$ evidence for the $\Xi_b^{0*}$ \cite{cms:xib}, a $bsu$ state of J=3/2. The state is reconstructed via the decay chain: $\Xi_b^{0*} \rightarrow \Xi_b^{-} \pi^+ $ with $\Xi_b^{-} \rightarrow J/\psi \Xi^-$, $J/\psi \rightarrow \mu \mu$, $\Xi^- \rightarrow \Lambda^0 \pi^-$ and $\Lambda^0 \rightarrow p \pi^-$. The mass of the  $\Xi_b^{0*}$ is found to be $5945.0 \pm 0.7~(stat.)~\pm 0.3~(syst.)~\pm 2.7 (PDG) $ MeV. The mass peak is visible in Fig. \ref{fig:e} (left).

CDF has produced the world's most precise measurements of the masses and widths of the $\Sigma_b^{\pm}$ and $\Sigma_b^{\pm *}$ \cite{cdf:sigmas}, two $bud$, S-wave isospin triplets with $J^P = 1/2^+ $ and $J^P = 3/2^+$, respectively.

\subsection{New bottomonium states}

\begin{figure}
\begin{center} 
    \includegraphics[width=0.45\textwidth]{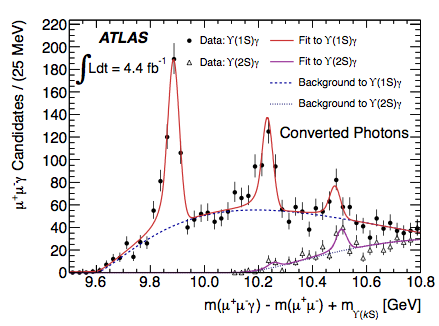}
    \includegraphics[width=0.45\textwidth]{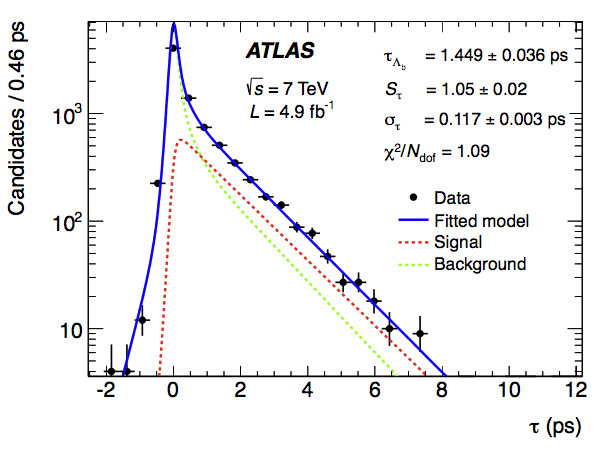}
\caption[*]{Left: the $\chi_b$(3P) signal discovered by ATLAS~\cite{atlas:chib3p}. Right: the lifetime of the $\Lambda_b$ in the paper by ATLAS~\cite{atlas:lambdab}.}
\label{fig:f}
\end{center}
\end{figure}

The ATLAS Collaboration has announced the discovery of a new resonance in the bottomonium region \cite{atlas:chib3p}, which is interpreted as the third radial excitation of the $^3P_J$ triplet, with $J=0,1,2$. The experimental resolution is not sufficient to resolve the three states, which are seen as a single peak. The 3P states are observed via their radiative decays $\chi_b(3P) \rightarrow \Upsilon(nS) + \gamma$, with n = 1,2, as can be seen in Fig. \ref{fig:f} (left). The barycenter of the triplet is reported at 10,530 $\pm$ 5 (stat.) $\pm$ 9 (syst.) MeV. The discovery is confirmed by D0 \cite{d0:chib3p} and LHCb \cite{lhcb:chib3p}. The observed position of the barycenter of the three states is in agreement with theoretical predictions. 

\section{Lifetimes}

The study of the lifetimes of hadrons containing a heavy quark gives important informations about the interplay between weak and strong interactions. In the so-called spectator model~\cite{spect} of HF decay, the lifetime of all b hadrons is the same, driven by the weak decay of the b quark. In reality, the strong interactions modify this scenario, leading to differences in the lifetimes of the b hadrons. These are predicted in the framework of the heavy-quark expansion (HQE) \cite{hqe}:

\[
\Gamma_B \sim |V_{CKM}|^2 \sum_n c_n(\mu) \left(  \frac{1}{m_b} \right)^n  <H_b|O_n|H_b>  .
 \]
The Wilson coefficients $c_n(\mu)$ contain the short-distance effects, and can be evaluated in perturbation theory. Long-distance physics is represented by the matrix element  $<H_b|O_n|H_b>$  that can be computed through nonperturbative QCD sum rules, operator-product-expansion methods or lattice QCD calculations. $V_{CKM}$ is the relevant Cabibbo-Kobayashi-Maskawa matrix element. In HQE, the order $1/m_b^2$ term distinguishes meson from baryon decays, while spectator effects of order $1/m_b^3$ differentiate between the lifetimes of $B^0$, $B^+$ and $B_s^0$ mesons.
For example, HQE predicts $\frac{\tau(B^0_s)}{\tau(B_0)} = 1.00 \pm 0.01$. 

The ATLAS Collaboration has presented a new measurement of the lifetime of the $\Lambda_b$ \cite{atlas:lambdab}, see Fig. \ref{fig:e} (right). This quantity has been controversial in past years  because of inconsistencies between the measurements. The ATLAS result is $\tau(\Lambda_b) = 1.499 \pm 0.036 \pm 0.017$ ps.

Finally, the lifetime of the $B_s$ is an extremely interesting quantity. Because of mixing, there are two mass eigenstates, $m_L$ and $m_H$, with separate lifetimes and widths. The LHCb Collaboration was able to establish that $m_H$ corresponds to the state with the longer lifetime \cite{lhcb:bs}. The final state in  $B_s \rightarrow J/\psi f_0(980)$ is CP odd, and allows the measurement of $\Gamma_H$, which was found to be $0.588 \pm 0.017 $ ps$^{-1}$. On the other hand, the channel $B_s \rightarrow K^+K^-$ is CP even, and, assuming CP is not violated, led to a value of   $\Gamma_L = 0.681 \pm 0.021 $ ps$^{-1}$. More details are given in the talk by Stephanie Hansmann-Menzemer.

\section{Conclusions}

In 2012, an impressive number of results concerning the production, spectroscopy and lifetimes of heavy-flavored hadrons has become available. Much of this was possible thanks to the LHC, but important contributions also came from experiments which are now shut down. 
Concerning HF production, cross section measurements of HF hadrons were presented in exclusive and inclusive decay modes, in association with jets, in association with vector bosons, and in other modes that were not possible to report here.  In the field of spectroscopy, the year saw the discovery of new mesons ($B^+_1$, $B^{+*}_2$ and $\chi_b(3P)$ ) and baryons ($\Lambda_b^*$ and $\Xi_b^*$). The lifetime of the $\Lambda_b$ was measured with improved precision, and the lifetime of the $B_s$ was studied in detail.

\end{document}